\documentclass[aps,pra,showpacs,groupedaddress,floatfix,nofootinbib]{revtex4}

\usepackage{amsmath}
\usepackage{graphicx}
\usepackage{color}
\usepackage{dcolumn}

\begin{document}

\title{Comment on ``Eigenstate clustering around exceptional points''}

\author{Francisco M. Fern\'andez}

\affiliation{INIFTA, Blvd. 113 y 64 S/N, Sucursal 4, Casilla de
Correo 16, 1900 La Plata, Argentina}

\begin{abstract}
We show that the author of a recent paper put forward some false statements
about the eigenstates of Hermitian and non-Hermitian systems. We conjecture
that one of the non-Hermitian Hamiltonians for a one-dimensional lattice is
similar to an Hermitian one and, consequently, exhibits real eigenvalues.
Present theoretical analysis of the eigenvalue equation suggests that one of
the sets of numerical results in the criticized paper may not be correct.
\end{abstract}


\maketitle

In a paper published recently in this journal Yuce\cite{Y20} proposed an
approach to identify eigenstate clustering in non-Hermitian
quantum-mechanical systems. Although the author states that ``Here, our aim
is to find the general condition of such closeness of eigenstates for a
given non-Hermitian Hamiltonian.'' he only showed some numerical results
derived from simple toy models. In this Comment we analyze Yuce's results
and some of his statements and remarks.

To begin with, we point out the following utterly false statement: ``In
Hermitian systems, all eigenstates are linearly dependent from each other
and hence no eigenstate clustering occurs.'' It is well known that different
eigenstates must be linearly independent, otherwise they are the same state
from a physical point of view. A second false statement is: ``As opposed to
the orthogonal eigenstates, nonorthogonal eigenstates are not linearly
dependent.'' Everybody knows that orthogonal vectors are \textit{always}
linearly independent while non-orthogonal vectors may be linearly dependent.

Another false statement is: ``In Hermitian systems, fidelities between any
two distinct eigenstates are always zero.'' Although the falsity of this
statement is obvious, in what follows we illustrate it by means of a simple
toy model. Consider the following matrix representation of an Hermitian
Hamiltonian operator in a vector space of dimension $3$:
\begin{equation}
\mathbf{H}=\left(
\begin{array}{lll}
0 & 1 & 1 \\
1 & 0 & 1 \\
1 & 1 & 0
\end{array}
\right) .  \label{eq:H_3x3}
\end{equation}
It has two eigenvalues $E_{1}=-1$ and $E_{2}=2$, the former two-fold
degenerate. The unnormalized column vectors $\mathbf{v}_{1}=\left(
\begin{array}{lll}
1 & 1 & -2
\end{array}
\right) ^{t}$ and $\mathbf{v}_{2}=\left(
\begin{array}{lll}
1+\xi & 1 & -2-\xi
\end{array}
\right) ^{t}$, where $t$ stands for transpose and $\xi $ is real, are
linearly independent, are eigenvectors of $\mathbf{H}$ with eigenvalue $%
E_{1} $ and have fidelity (see Yuce's Eq.~(2) for its definition)
\begin{equation}
F_{12}=\frac{3\left( \xi +2\right) ^{2}}{4\left( \xi ^{2}+3\xi +3\right) }.
\label{eq:F12}
\end{equation}
When $\xi =0$, then $F_{12}=1$ and the two vectors are linearly dependent
(the same eigenstate from a physical point of view), otherwise they are
linearly independent. In particular, when $\xi =-2$ they are orthogonal. The
correct expression would be: the eigenstates of an Hermitian operator
\textit{can always be chosen} to be orthogonal and, consequently, with zero
fidelities. This may not be possible in the case of a non-Hermitian operator
unless one resorts to a suitable metric\cite{Z10} (and references therein).

As an illustrative example Yuce selected an $N$-level non-Hermitian
Hamiltonian with eigenstates $\psi _{j}$. More precisely, it is a
one-dimensional lattice with $N$ sites having gain or loss impurities and
site-dependent nearest-neighboring hopping amplitudes $t_{n}$ and $%
t_{n}^{\prime }$ in the forward and backward directions, respectively. As a
realization of this model he wrote: $\mathcal{H}\psi _{n}=t_{n}\psi
_{n+1}+t_{n-1}^{\prime }\psi _{n-1}+i\gamma _{n}\psi _{n}$. It is clear that
the $\psi _{n}$ in this equation are not the eigenstates mentioned
previously. If $\mathcal{H}$ stands for the Hamiltonian operator, then the $%
\psi _{n}$ should be some kind of vectors. In his figures Yuce plotted
densities $\left| \psi _{(j)}\right| ^{2}$ which do not seem to be properly
defined in the paper but they are probably related to the $\psi _{n}$ of the
equation just mentioned. Therefore, in order to analyze Yuce's results we
will try some assumptions and guesses.

We assume that the Hamiltonian operator for the one-dimensional lattice is
given by
\begin{equation}
H=\sum_{j=1}^{N-1}\left( t_{j}\left| j\right\rangle \left\langle j+1\right|
+t_{j}^{\prime }\left| j+1\right\rangle \left\langle j\right| \right)
+i\sum_{j=1}^{N}\gamma _{j}\left| j\right\rangle \left\langle j\right| ,
\label{eq:H_lattice}
\end{equation}
where $\left\{ \left| j\right\rangle ,\;j=1,2,\ldots ,N\right\} $ is an
orthonormal basis set. If $\gamma _{j}=0$ and $t_{j}^{\prime }=t_{n}^{*}$
this Hamiltonian operator is Hermitian. Its eigenvectors can be written in
terms of the basis vectors as
\begin{equation}
\left| \psi \right\rangle =\sum_{j=1}^{N}\psi _{j}\left| j\right\rangle .
\label{eq:|psi>}
\end{equation}
If we insert this expression into $H\left| \psi \right\rangle =E\left| \psi
\right\rangle $, then the coefficients $\psi _{j}$ satisfy the three-term
recurrence relation
\begin{eqnarray}
t_{n-1}^{\prime }\psi _{n-1}+\left( i\gamma _{n}-E\right) \psi
_{n}+t_{n}\psi _{n+1} &=&0,\;n=1,2,\ldots ,N,  \nonumber \\
\psi _{0} &=&\psi _{N+1}=0.  \label{eq:secular_1}
\end{eqnarray}
Note that the boundary conditions apply to a linear chain (with open ends).
For every eigenvalue $E_{k}$, $k=1,2,\ldots ,M\leq N$, we have an
eigenvector $\left| \psi _{k}\right\rangle $ with coefficients $\psi _{jk}$
and we assume that the densities plotted by Yuce are given by $\left| \psi
_{jk}\right| ^{2}$ (we have one curve for every value of $k$). Comparison of
these curves makes sense if all the eigenvectors are normalized in the same
way; for example, $\left\langle \psi _{k}\right| \left. \psi
_{k}\right\rangle =1$. In his Fig.1 Yuce chose $\gamma _{n}=0$ and $t_{n}$, $%
t_{n}^{\prime }$ real, so that in what follows we consider this particular
situation.

In order to analyze the secular equation (\ref{eq:secular_1}) we resort to
an argument used earlier with the purpose of truncating three-term
recurrence relations\cite{CDW00,AF20}. If we substitute $\psi _{n}=Q_{n}c_{n}
$ into Eq.~(\ref{eq:secular_1}) and divide the resulting expression by $Q_{n}
$ we obtain
\begin{equation}
t_{n-1}^{\prime }\frac{Q_{n-1}}{Q_{n}}c_{n-1}-Ec_{n}+t_{n}\frac{Q_{n+1}}{%
Q_{n}}c_{n+1}=0.  \label{eq:secular_Q}
\end{equation}
The matrix representation that leads to this secular equation is symmetric
if $t_{n}Q_{n+1}^{2}=t_{n}^{\prime }Q_{n}^{2}$. Therefore, we conclude that
if $\gamma _{n}=0$ and $t_{n}t_{n}^{\prime }>0$ then the Hamiltonian
operator (\ref{eq:H_lattice}) is similar to an Hermitian Hamiltonian and its
eigenvalues are real. On choosing $Q_{n+1}/Q_{n}=t_{n}^{\prime }/t_{n}$ the
secular equation (\ref{eq:secular_Q}) becomes
\begin{equation}
\sqrt{t_{n-1}t_{n-1}^{\prime }}c_{n-1}-Ec_{n}+\sqrt{t_{n}t_{n}^{\prime }}%
c_{n+1}=0.  \label{eq:secular_symmetric}
\end{equation}
This argument applies to the open chain because the boundary conditions
remain unchanged: $c_{0}=c_{N+1}=0$. On the other hand, it does not apply to
a closed chain with boundary conditions $\psi _{0}=\psi _{N}$ because $%
c_{0}\neq c_{N}$.

The argument above fails when one or more pairs of model parameters satisfy $%
t_{n}t_{n}^{\prime }<0$ so that we expect exceptional points when some $%
t_{n} $ or $t_{n}^{\prime }$ vanish. For this reason, we can
obtain exceptional points of any order by tuning the model
parameters. Close to those exceptional points we expect the
eigenstate clustering studied by Yuce. It is not our purpose to
discuss this feature in detail here.

In what follows we obtain another useful result about the eigenvalues and
eigenvectors of the secular equation (\ref{eq:secular_1}) with $\gamma
_{n}=0 $. One can easily verify that if $\psi _{n}$, $n=1,2,\ldots ,N$, is a
solution to the secular equation (\ref{eq:secular_1}) with $\gamma _{n}=0$
for some value of $E$ then $(-1)^{n}\psi _{n}$ is also a solution for $-E$;
in other words, we can always obtain pairs of eigenvectors by means of the
relation: $\psi _{n}(-E)=(-1)^{n}\psi _{n}(E)$. Note that both are suitable
solutions because $\psi _{0}(\pm E)=\psi _{N+1}(\pm E)=0$. When $N$ is odd
there is always a solution with energy $E=0$. Another conclusion is that if
the eigenvectors are normalized in the same way we expect $N/2$ different
curves $\left| \psi _{jk}\right| ^{2}$ for $N$ even and $(N+1)/2$ when $N$
is odd. However, Yuce's Fig.~1~(a) for $N=12$, $t_{n}=2t_{n}^{\prime }=0.1$
and open boundary conditions shows more than $6$ curves. Therefore, either
his results are wrong or his eigenvectors are not properly normalized or we
are mistaken about what it was plotted in those figures. Unfortunately,
Yuce's extremely unclear notation makes it difficult a more thorough
analysis of his results and conclusions. It is worth mentioning that if $%
t_{n}=t$ and $t_{n}^{\prime }=t^{\prime }$ are both constant, and $%
tt^{\prime }>0$, then the eigenvalue equation is exactly solvable.

Summarizing: in this comment we have shown that at least three of Yuce's
statements are false and our theoretical analysis suggests that some
numerical results appear to be wrong.

\end{document}